\documentclass{article}

\usepackage[preprint]{neurips_2023}

\usepackage[utf8]{inputenc} 
\usepackage[T1]{fontenc}    
\usepackage{hyperref}       
\usepackage{url}            
\usepackage{booktabs}       
\usepackage{amsfonts}       
\usepackage{nicefrac}       
\usepackage{microtype}      
\usepackage{xcolor}         

\usepackage{graphicx}
\usepackage{natbib}
\setcitestyle{numbers,square}
\usepackage{doi}
\usepackage{bm}
\usepackage{amsmath,amsthm,amssymb}
\usepackage[linesnumbered,ruled]{algorithm2e}
\usepackage{wrapfig}
\usepackage[multiple]{footmisc}

\newtheorem{definition}{Definition}

\newtheorem{theorem}{Theorem}
\newtheorem{lemma}{Lemma}

\title{Score-Based Equilibrium Learning in Multi-Player Finite Games with Imperfect Information}

\author{%
  Runyu Lu\textnormal{\textsuperscript{1}}\\
  \texttt{lurunyu17@mails.ucas.ac.cn}\\
  \And
  Yuanheng Zhu\textnormal{\textsuperscript{2}}\\
  \texttt{yuanheng.zhu@ia.ac.cn}\\
  \And
  Dongbin Zhao\textnormal{\textsuperscript{3}}\\
  \texttt{dongbin.zhao@ia.ac.cn}
}

\begin{document}

\maketitle

\footnotetext[1]{School of Artificial Intelligence, University of Chinese Academy of Sciences \textsuperscript{2,3}State Key Laboratory of Multimodal Artificial Intelligence Systems, Institute of Automation, Chinese Academy of Sciences}

\begin{abstract}
  Real-world games, which concern imperfect information, multiple players, and simultaneous moves, are less frequently discussed in the existing literature of game theory. While reinforcement learning (RL) provides a general framework to extend the game theoretical algorithms, the assumptions that guarantee their convergence towards Nash equilibria may no longer hold in real-world games. Starting from the definition of the Nash distribution, we construct a continuous-time dynamic named imperfect-information exponential-decay score-based learning (IESL) to find approximate Nash equilibria in games with the above-mentioned features. Theoretical analysis demonstrates that IESL yields equilibrium-approaching policies in imperfect information simultaneous games with the basic assumption of concavity. Experimental results show that IESL manages to find approximate Nash equilibria in four canonical poker scenarios and significantly outperforms three other representative algorithms in 3-player Leduc poker, manifesting its equilibrium-finding ability even in practical sequential games. Furthermore, related to the concept of game hypomonotonicity, a trade-off between the convergence of the IESL dynamic and the ultimate NashConv of the convergent policies is observed from the perspectives of both theory and experiment.
\end{abstract}

\section{Introduction}

\subsection{Background}

Allowing players to bluff and probe, imperfect information games have become an interesting topic in recent research involving both game theory and machine learning. This class of games covers more real-life problems like poker (see \cite{moravvcik2017deepstack}\cite{brown2018superhuman}) and is even more difficult to solve when compared with its perfect information counterpart. Despite its complexity, research in game theory (see \cite{roughgarden2016twenty}\cite{nisan2007algorithmic}) has established various equilibrium-finding dynamics that have shown the potential to solve large-scale imperfect information games with the help of deep reinforcement learning (DRL).

As two typical equilibrium-learning dynamics, fictitious play (FP) \cite{brown1951iterative} and regret matching \cite{hart2000simple} (RM) yield average policies that converge to Nash equilibria in two-player zero-sum games (see \cite{robinson1951iterative}\cite{waugh2009abstraction}). In sequential imperfect information games, they can be extended into extensive-form fictitious play (XFP) \cite{heinrich2015fictitious} and counterfactual regret minimization (CFR) \cite{zinkevich2007regret}, respectively. Using DRL techniques, Neural Fictitious Self-Play (NFSP) \cite{heinrich2016deep} and Policy-Space Response Oracles (PSRO) \cite{lanctot2017unified} further extend FP to large-scale games, while Deep CFR \cite{brown2019deep} extends Monte Carlo CFR \cite{lanctot2013monte} in a similar manner. As a score-based learning dynamic that originates from evolutionary game theory \cite{zeeman2006population}\cite{hofbauer1998evolutionary}, Follow the Regularized Leader (FTRL) \cite{mcmahan2011follow} generalizes replicator dynamics (RD) \cite{taylor1978evolutionary} and guarantees average-policy convergence to Nash equilibrium also in zero-sum games. NeuRD \cite{hennes2020neural} further extends RD with policy-based reinforcement learning and into imperfect information games.

\subsection{Recent Works}

Although the methods above are capable of finding Nash equilibria in zero-sum scenarios, the preservation of average policies can be difficult when function approximators like neural networks are introduced. Perolat et al. \cite{perolat2021poincare} showed that the \textit{instantaneous policy} in FTRL may cycle in imperfect information games and further proposed an algorithm where last-iterate convergence towards exact Nash equilibrium is guaranteed in zero-sum or monotone games. Borrowing the one-step policy-update idea from NeuRD, the algorithm named Regularized Nash Dynamic (R-NaD) was combined with DRL to build DeepNash \cite{perolat2022mastering}, which achieved human expert-level performance in Stratego.

While R-NaD applies reward regularization to ensure the convergence of instantaneous policies, another route is to study a broader class of score-based learning dynamics. Gao et al. \cite{gao2020passivity} showed that a modified version of the score function, whose integral expression contains an exponentially decaying term, induces instantaneous policies that approach Nash distributions in normal-form games characterized by hypomonotonicity. The corresponding dynamic EXP-D-RL was generalized into Markov games by Zhu et al. \cite{zhu2022empirical}. In this paper, by proposing the continuous-time dynamic named \textit{imperfect-information exponential-decay score-based learning} (IESL), we will show that a proper score construction can give rise to direct policy convergence towards approximate Nash equilibrium even in multi-player imperfect information games, both theoretically and empirically.

\subsection{Motivations \& Contributions}

While the above-mentioned methods for dealing with imperfect information heavily rely on the formulation of extensive-form games, which require players to make decisions in turn, many real-world games (even real-time video games) allow players to move simultaneously. Since reinforcement learning is primarily established upon Markov decision processes (MDPs), which are characterized by simultaneous moves, building learning dynamics with the language of RL naturally deals with this problem and allows further extensions with theoretical guarantees. While partially observable stochastic games (POSGs) \cite{hansen2004dynamic}) directly extend MDPs into imperfect information games, existing research on POSGs usually requires additional strong assumptions when constructing equilibrium-learning algorithms (see PO-POSG \cite{horak2019solving}). In this paper, we restate POSG as a general framework to describe imperfect information simultaneous games and show that IESL yields approximate Nash equilibria with a minimum of further restrictions on the game. While our theoretical analysis is based on this POSG-like framework, IESL can be rightly implemented in sequential games like poker and practically compared with algorithms like CFR.

The contributions of this paper include:

\begin{itemize}
\item We theoretically analyze the relationships among two variants of Nash distribution, the rest point in IESL, and approximate Nash equilibrium in finite games with imperfect information and simultaneous moves. An important property is that every rest point of IESL induces an approximate Nash equilibrium when the utility is concave.

\item We theoretically demonstrate that IESL rest points attract other points in nearby regions restricted by policy hypomonotonicity. Since we have proved that the rest points of IESL yield approximate Nash equilibria, a novel convergence result with regard to Nash equilibrium in multi-player imperfect information games is established.

\item We empirically verify that the instantaneous policies in IESL do converge to approximate Nash equilibria in multi-player poker games. In fact, IESL significantly outperforms CFR, FP, and RD in moderate-scale 3-player Leduc poker.

\item From both theory and experiment, we observe that a trade-off exists between the convergence of the IESL dynamic and the NashConv of the convergent policies. This induces an efficient parameter selection method when we seek the lowest NashConv of a convergent policy.
\end{itemize}

\section{Preliminaries}

\subsection{Game Description \& Useful Notations}

To describe general finite games with imperfect information and simultaneous moves, we first restate POSG under the restriction of a finite horizon. (See Appendix A.2 for a more detailed description.)

\begin{definition}
An $n$-player finite-horizon imperfect information game with simultaneous moves is a tuple $\left\langle N,S,Z,A,Tr,O,\left\{{{R}^{i}}\right\},\rho,\gamma,T\right\rangle$. $N=\{1,2,\cdots,n\}$ is the set of players. $S$ is the set of global states $s$. $Z$ is the set of joint observations with $\vec{z}={(z^i)}_{i\in N}\in Z$. $A$ is the set of actions with $\vec{a}={(a^i)}_{i\in N}\in A^n$. $Tr$ is the tuple of state transition distributions that generate subsequent states with $s_{t+1}\sim Tr(s_t,\vec{a}_t)$. $O$ is the tuple of joint observation distributions that generates observations with $\vec{z}_{t+1}\sim O(s_t,\vec{a}_t,s_{t+1})$. $R^i$ is the tuple of reward functions that generate player $i$'s instant rewards with $r_{t+1}^{i}= R^i(s_t,\vec{a}_t,s_{t+1})$. $\rho$ is the initial state distribution that generates the initial state $s_0$ with $s_0\sim \rho$. $\gamma\in(0,1]$ is the discount factor. $T$ is the termination time.
\end{definition}

In addition to the basic elements in the definition above, some extra notations are useful in analysis.

Borrowing an important idea from extensive-form games, we can use trajectories to describe an arbitrary situation in a POSG. A global trajectory at step $t$ $(t<T)$ can be expressed as a history $h_t=\left\{s_0,\vec{z}_0,\vec{a}_0,\cdots,s_{t-1},\vec{z}_{t-1},\vec{a}_{t-1},s_t,\vec{z}_t\right\}$. In comparison, a local trajectory for player $i$ can be expressed as a reduced version $x_t^{i}=\left\{z_0^i,a_0^i,\cdots,z_{t-1}^i,a_{t-1}^i,z_t^i\right\}$ that corresponds to an information set defined over all possible $h_{t}$, as in extensive-form games. We write $h_t\in x_t^i$ if $h_t$ belongs to the corresponding information set of $x_t^i$, or, equivalently, if $x_t^i$ is compatible with $h_t$ in reduction. For $k\geq t$, we simply write $h_k\sqsupset h_t$ if $h_t$ is compatible with $h_k$, and $x_k^i\sqsupset x_t^i$ if $x_t^i$ is compatible with $x_k^i$.

Policy is the core concept when we analyze a single-player decision process or a multiple-player game. In an imperfect information simultaneous game, player $i$'s policy $\pi^i$ at each local trajectory (information set) $x^i$ is a probability simplex over $A$, assigning each possible action $a^i$ a probability $\pi^i(x^i,a^i)$. The set of all complete policies for player $i$ is expressed as $\Pi^i$. We use the joint policy $\vec{\pi}=(\pi^i)_{i\in N}$ to denote a combination of all players' policies, and use $\vec{\pi}^{-i}$ to denote a combination of all players' policies except player $i$.

Following single-agent Markov decision processes, we can define global-trajectory-based value functions $V_{{\vec{\pi }}}^{i}({{h}_{t}})=\mathbb{E}\left[ \sum\nolimits_{k=t}^{T-1}{{{\gamma }^{k-t}}r_{k+1}^{i}}\left| {{h}_{t}},\vec{\pi } \right. \right]$ and $Q_{{\vec{\pi }}}^{i}({{h}_{t}},a_{t}^{i})=\mathbb{E}\left[ r_{t+1}^{i}+\gamma V_{{\vec{\pi }}}^{i}({{h}_{t+1}})\left| {{h}_{t}},a_{t}^{i},\vec{\pi } \right. \right]$, as well as the advantage function $A_{{\vec{\pi }}}^{i}({{h}_{t}},a_{t}^{i})=Q_{{\vec{\pi }}}^{i}({{h}_{t}},a_{t}^{i})-V_{{\vec{\pi }}}^{i}({{h}_{t}})$.

\subsection{Nash Equilibrium \& Nash Distribution}

By constructing utility functions in different forms, various rest points in policy space can be defined. Nash equilibrium is among the most well-studied classes of rest points, reflecting a stable condition with respect to individual payoff maximization.

\begin{definition}
A Nash equilibrium $\vec{\pi}_{nash}$ in an $n$-player finite-horizon imperfect information game with simultaneous moves is described using the utility function: ${{u}^{i}}(\vec{\pi })={{\mathbb{E}}_{{{h}_{0}}}}\left[ V_{{\vec{\pi }}}^{i}({{h}_{0}}) \right]$. For any $i\in N$ and for any possible policy $\pi^i\in\Pi^i$, it holds:
\begin{align*}
{{u}^{i}}({{\vec{\pi }}_{nash}})\ge {{u}^{i}}({{\pi }^{i}},\vec{\pi }_{nash}^{-i})
\end{align*}
\end{definition}

NashConv is a commonly used concept that measures the closeness of an arbitrary policy $\vec{\pi}$ towards Nash equilibrium: $\mathrm{NashConv}(\vec{\pi})=\sum\nolimits_{i=1}^{n}{{{\max }_{\pi _{\dagger }^{i}\in {{\Pi }^{i}}}}\left\{ {{u}^{i}}(\pi _{\dagger }^{i},{\vec{\pi }^{-i}}) \right\}-{{u}^{i}}(\vec{\pi })}$. Exact Nash equilibrium has zero NashConv, and its computation in non-zero-sum games is probably rather hard (see \cite{etessami2010complexity}). Therefore, it is reasonable to consider approximate Nash equilibrium in multi-player games.

\begin{definition}
An $\epsilon^\prime$-Nash equilibrium $\vec{\pi}_{nash(\epsilon^\prime)}$ in an $n$-player finite-horizon imperfect information game with simultaneous moves is described using the utility function: ${{u}^{i}}(\vec{\pi })={{\mathbb{E}}_{{{h}_{0}}}}\left[ V_{{\vec{\pi }}}^{i}({{h}_{0}}) \right]$. For any $i\in N$ and for any possible policy $\pi^i\in\Pi^i$, it holds:
\begin{align*}
{{u}^{i}}({{\pi }^{i}},\vec{\pi }_{nash(\epsilon^\prime)}^{-i})-{{u}^{i}}({{\vec{\pi }}_{nash(\epsilon^\prime)}})\leq\epsilon^\prime
\end{align*}
\end{definition}

In the following paragraphs, we refer to the infimum of $\epsilon^\prime$ as \textit{deviation}. Simply by definition, we know that an approximate Nash equilibrium has NashConv upper-bounded by the deviation multiplied by $n$. As finding even approximate Nash equilibrium is proved to be PPAD-complete \cite{daskalakis2009complexity}, constructing dynamics directly related to this concept is still non-trivial. Nash distribution \cite{leslie2005individual} describes another rest point class with utility regularization, which is helpful in building convergent learning dynamics.

\begin{definition}
A Nash distribution $\vec{\pi}_{*}$ characterized by a non-negative parameter $\epsilon$ in an $n$-player finite-horizon imperfect information game with simultaneous moves is described using the regularized utility function $\tilde{u}^{i}(\vec{\pi })={{\mathbb{E}}_{{{h}_{0}}}}\left[ V_{{\vec{\pi }}}^{i}({{h}_{0}}) \right]-\epsilon \sum\limits_{t=0}^{T-1}{{{\gamma }^{t}}\sum\limits_{x_{t}^{i}}{\left( H({{\pi }^{i}}(x_{t}^{i}))\sum\limits_{{{h}_{t}}\in x_{t}^{i}}{\Pr \left( {{h}_{t}}|\vec{\pi } \right)} \right)}}$, where $H({{\pi }^{i}}(x_{t}^{i}))=\sum\limits_{a_{t}^{i}}{{{\pi }^{i}}(x_{t}^{i},a_{t}^{i})\log {{\pi }^{i}}(x_{t}^{i},a_{t}^{i})}$ is the negative Gibbs entropy. For any $i\in N$ and for any possible policy $\pi^i\in\Pi^i$, it holds:
\begin{align}
{\tilde{u}^{i}}({{\vec{\pi }}_{*}})\ge {\tilde{u}^{i}}({{\pi }^{i}},\vec{\pi }_{*}^{-i})
\label{Nash Distribution}
\end{align}
\end{definition}

In \textbf{Definition 4}, the expression of $\tilde{u}^{i}(\vec{\pi })$ extracts the same expected value term ${{\mathbb{E}}_{{{h}_{0}}}}\left[ V_{{\vec{\pi }}}^{i}({{h}_{0}}) \right]$ in comparison to Nash equilibrium. Alternatively, we can write $\tilde{u}^{i}(\vec{\pi })={{\mathbb{E}}_{{{h}_{0}}}}\left[ \tilde{V}_{{\vec{\pi }}}^{i}({{h}_{0}}) \right]$, where $\tilde{V}$ is the value defined under a regularized reward function $\tilde{R}^{x_t^i}(s_t,\vec{a}_t,s_{t+1})=R^i(s_t,\vec{a}_t,s_{t+1})-H({{\pi }^{i}}(x_{t}^{i}))$. This defines the same utility function as in the formal definition above.

Apparently, when we set $\epsilon^\prime=0$ for approximate Nash equilibrium and $\epsilon=0$ for Nash distribution, the two concepts become the same one that describes exact Nash equilibrium. However, because the expression of $\tilde{u}^{i}(\vec{\pi })$ contains a reaching probability $\Pr \left( {{h}_{t}}|\vec{\pi } \right)$ determined by the own policy $\vec{\pi}$, the deviation of a Nash distribution is not tightly bounded. Besides, the probability term makes it hard to prove the relationship between a Nash distribution and the rest point of a learning dynamic in an imperfect information game. To make practical use of this concept, we consider modifying the original definition of Nash distribution. In Section 3, two variants of the Nash distribution are proposed, and they eventually serve to construct our equilibrium-finding dynamic.

\section{Dynamic Construction}

\subsection{Variants of Nash Distribution}

\begin{definition}
Characterized by a non-negative parameter $\epsilon$, a Nash distribution with \textbf{modification-L} is a policy $\vec{\pi}_L$. For any $i\in N$ and for any possible policy $\pi^i\in\Pi^i$, it holds:
\begin{align*}
&{{u}^{i}}({{\vec{\pi }}_{L}})-\epsilon\sum\limits_{t=0}^{T-1}{{{\gamma }^{t}}\sum\limits_{x_{t}^{i}}{\left( H(\pi _{L}^{i}(x_{t}^{i}))\sum\limits_{{{h}_{t}}\in x_{t}^{i}}{\Pr \left( {{h}_{t}}|{{{\vec{\pi }}}_{L}} \right)} \right)}}\ge\\
&{{u}^{i}}({{\pi }^{i}},\vec{\pi }_{L}^{-i})-\epsilon\sum\limits_{t=0}^{T-1}{{{\gamma }^{t}}\sum\limits_{x_{t}^{i}}{\left( H({{\pi }^{i}}(x_{t}^{i}))\sum\limits_{{{h}_{t}}\in x_{t}^{i}}{\Pr \left( {{h}_{t}}|{{{\vec{\pi }}}_{L}} \right)} \right)}}
\end{align*}
where ${{u}^{i}}(\vec{\pi })={{\mathbb{E}}_{{{h}_{0}}}}\left[ V_{{\vec{\pi }}}^{i}({{h}_{0}}) \right]$.
\end{definition}

Note that \textbf{modification-L} replaces the RHS probability $\Pr \left( {{h}_{t}}|{\pi^i,{\vec{\pi }^{-i}_L}} \right)$ in the original definition of Nash distribution (\ref{Nash Distribution}) with the same $\Pr \left( {{h}_{t}}|\vec{\pi}_L \right)$ term in the LHS. Through transposition, we have:
\begin{align*}
{{u}^{i}}({{\pi }^{i}},\vec{\pi }_{L}^{-i})-{{u}^{i}}({{\vec{\pi }}_{L}})\le \epsilon \sum\limits_{t=0}^{T-1}{{{\gamma }^{t}}\sum\limits_{x_{t}^{i}}{\left( \left( H({{\pi }^{i}}(x_{t}^{i}))-H(\pi _{L}^{i}(x_{t}^{i})) \right)\sum\limits_{{{h}_{t}}\in x_{t}^{i}}{\Pr \left( {{h}_{t}}|{{{\vec{\pi }}}_{L}} \right)} \right)}}
\end{align*}

Since $H(\cdot)$ is the negative Gibbs entropy, $\left( H({{\pi }^{i}}(x_{t}^{i}))-H(\pi _{L}^{i}(x_{t}^{i})) \right)$ has an upper bound $\log{|A|}$. Since the probability $\Pr \left( {{h}_{t}}|\vec{\pi}_L \right)$ sums up to $1$ over all global trajectories $h_t$ with the same $t$, the deviation of $\vec{\pi}_L$ is bounded by $\epsilon \log \left| A \right|\sum\nolimits_{k=0}^{T-1}{{{\gamma }^{k}}}$.

\begin{definition}
Characterized by a non-negative parameter $\epsilon$, a Nash distribution with \textbf{modification-R} is a policy $\vec{\pi}_R$. For any $i\in N$ and for any possible policy $\pi^i\in\Pi^i$, it holds:
\begin{align*}
&{{u}^{i}}({{\vec{\pi }}_{R}})-\epsilon\sum\limits_{t=0}^{T-1}{{{\gamma }^{t}}\sum\limits_{x_{t}^{i}}{\left( H(\pi _{R}^{i}(x_{t}^{i}))\sum\limits_{{{h}_{t}}\in x_{t}^{i}}{\Pr \left( {{h}_{t}}|{\pi^i,{\vec{\pi }^{-i}_R}} \right)} \right)}}\ge\\
&{{u}^{i}}({{\pi }^{i}},\vec{\pi }_{R}^{-i})-\epsilon\sum\limits_{t=0}^{T-1}{{{\gamma }^{t}}\sum\limits_{x_{t}^{i}}{\left( H({{\pi }^{i}}(x_{t}^{i}))\sum\limits_{{{h}_{t}}\in x_{t}^{i}}{\Pr \left( {{h}_{t}}|{\pi^i,{\vec{\pi }^{-i}_R}} \right)} \right)}}
\end{align*}
where ${{u}^{i}}(\vec{\pi })={{\mathbb{E}}_{{{h}_{0}}}}\left[ V_{{\vec{\pi }}}^{i}({{h}_{0}}) \right]$.
\end{definition}

Note that \textbf{modification-R} replaces the LHS probability $\Pr \left( {{h}_{t}}|\vec{\pi }_R \right)$ in the original definition of Nash distribution (\ref{Nash Distribution}) with the same $\Pr \left( {{h}_{t}}|{\pi^i,{\vec{\pi }^{-i}_R}} \right)$ term in the RHS. Similarly, we can show that the deviation of $\vec{\pi}_R$ is bounded by $\epsilon \log \left| A \right|\sum\nolimits_{k=0}^{T-1}{{{\gamma }^{k}}}$, as a common reaching probability is used on both sides to construct the policy-entropy regularization terms.

The two variants of the Nash distribution guide us to build a related learning dynamic whose rest points also yield policies with a bounded deviation.

\subsection{Main Dynamic}

Our main dynamic, named imperfect-information exponential-decay score-based learning (IESL), is expressed with (\ref{IESL})(\ref{w})(\ref{sigma-def}). Note that both score $\vec{y}$ and policy $\vec{\pi}$ are functions on a continuous time variable $t$, and the dot over $y$ indicates its derivative on $t$. Since the game description has used the same variable as the discrete timestep, to avoid ambiguity, we will not explicitly write out the subscript $t$ in IESL unless necessary.
\begin{align}
\left\{ \begin{matrix}
\ \ {{{\dot{\vec{y}}}}}=\eta (\vec{w}({{{\vec{\pi }}}})-{{{\vec{y}}}})  \\\\
\ \ {{{\vec{\pi }}}}=\vec{\sigma }({{{\vec{y}}}})
\end{matrix} \right.
\label{IESL}
\end{align}

In comparison to replicator dynamics, where ${{{\dot{\vec{y}}}}}=\vec{w}({{{\vec{\pi }}}})$, IESL introduces an exponential decay in the integral form of the score:
\begin{align}
y_{t}^{i}({{x}^{i}},{{a}^{i}})={{e}^{-\eta t}}y_{0}^{i}({{x}^{i}},{{a}^{i}})+\eta \int_{0}^{t}{{{e}^{-\eta (t-\tau )}}{{w}^{i}}({{{\vec{\pi }}}_{\tau }})({{x}^{i}},{{a}^{i}})d\tau }
\label{IESL-integral}
\end{align}

In the context of imperfect information games, $\vec{w}(\cdot)$ in (\ref{IESL}) can be viewed as a vectored local advantage function from the perspective of each individual player. To be specific, $w^i(\cdot)$ averages globally defined advantage values over all histories $h^i$ that belong to a given information set $x^i$:
\begin{align}
{{w}^{i}}({{\vec{\pi }}})({{x}^{i}},{{a}^{i}})=\frac{\sum\limits_{{{h}^{i}}\in {{x}^{i}}}{\Pr \left( {{h}^{i}}|{{{\vec{\pi }}}} \right)A_{{{{\vec{\pi }}}}}^{i}({{h}^{i}},{{a}^{i}})}}{\sum\limits_{{{h}^{i}}\in {{x}^{i}}}{\Pr \left( {{h}^{i}}|{{{\vec{\pi }}}} \right)}}
\label{w}
\end{align}

The regularized policy choice function $\vec{\sigma}(\cdot)$ in (\ref{IESL}) maps the score $\vec{y}$ to a corresponding policy $\vec{\pi}$:
\begin{align}
{{\sigma }^{i}}(y^i)({{x}^{i}})=\underset{{{\pi }^{i}}({{x}^{i}})\in {{\Pi }^{i}}({{x}^{i}})}{\mathop{\arg \max }}\,\left\{ \sum\limits_{{a}^{i}}{{{\pi }^{i}}({{x}^{i}},{{a}^{i}})y^{i}({{x}^{i}},{{a}^{i}})}-\epsilon H({{\pi }^{i}}({{x}^{i}})) \right\}
\label{sigma-def}
\end{align}

Since $H$ is the negative Gibbs entropy, $\vec{\sigma}(\cdot)$ has a close-form expression, which is known as softmax:
\begin{align}
{{\sigma }^{i}}(y^i)({{x}^{i}},{{a}^{i}})=\frac{\exp \left( \frac{1}{\epsilon }{{y}^{i}}({{x}^{i}},{{a}^{i}}) \right)}{\sum\limits_{b}{\exp \left( \frac{1}{\epsilon }{{y}^{i}}({{x}^{i}},b) \right)}}
\label{sigma-softmax}
\end{align}

Note that it generates pure strategy with the largest score when $\epsilon\to0^+$ (fully exploit) and uniformly random policy when $\epsilon\to+\infty$ (fully explore). Also note that $\epsilon H(\cdot)$ is an $\epsilon$-strongly convex penalty function (see \cite{mertikopoulos2019learning} Section 3.2), which is essential in the subsequent convergence analysis of IESL.

We can see that in (\ref{IESL-integral}), the exponentially decaying factor leads to a provably bounded score. On the one hand, $\vec{w}(\cdot)$ is naturally bounded since the game is finite. On the other hand, $\int_{0}^{t}{{{e}^{-\eta (t-\tau )}}d\tau }=\frac{1}{\eta }\left. {{e}^{-\eta (t-\tau )}} \right|_{0}^{t}=\frac{1}{\eta }(1-{{e}^{-\eta t}})\in [0,\frac{1}{\eta }]$. The basic property of integrals shows that $\vec{y}$ is bounded.

Besides, since $\vec{\sigma}(\cdot)$ is continuous, $\vec{w}(\vec{\sigma}(\cdot))$ is also continuous. Therefore, a fixed point of $\vec{w}(\vec{\sigma}(\cdot))$ always exists by Brouwer's fixed-point theorem \cite{florenzano2003general}. The existence of a rest point of (\ref{IESL}) in score space is thus guaranteed. In Section 3.3, we will further show that any Nash distribution with either modification described in Section 3.1 induces a rest point with regard to IESL.

\subsection{Intrinsic Relationship}

As we have mentioned, the construction of the IESL dynamic is derived from the definitions of the two variants of the Nash distribution. Once we relate the two concepts to the rest point of our learning dynamic, the direct bound on their deviation may suggest a similar bound on that of the corresponding policy at a rest point in IESL.

A value discrepancy arising from a policy difference can be decomposed as a sum of per-timestep advantages (see \cite{schulman2015trust}). Since the property derived from single-agent Markov decision processes is quite useful in analysis, we first introduce it into multi-player imperfect information games.

\begin{lemma}
Suppose a policy $\pi^{i}_{\dagger}$ that differs from $\pi^i$ for player $i$. Then the policy difference yields:
\begin{align*}
V_{\pi _{\dagger }^{i},{{{\vec{\pi }}}^{-i}}}^{i}({{h}_{t}})-V_{{\vec{\pi }}}^{i}({{h}_{t}})=\mathbb{E}\left[ \sum\limits_{k=t}^{T-1}{{{\gamma }^{k-t}}A_{{\vec{\pi }}}^{i}({{h}_{k}},a_{k}^{i})}\left| {{h}_{t}},\pi _{\dagger }^{i},{{{\vec{\pi }}}^{-i}} \right. \right]
\end{align*}
\end{lemma}

Two proofs for \textbf{Lemma 1} are provided in Appendix B.1 \& B.2. The first proof is concise, while the second proof is more detailed by showing the exact process of state transition and reward generation with the basic elements in \textbf{Definition 1}.

Using the property of policy difference, we can prove an important inequality for both of the two variants of the Nash distribution.

\begin{lemma}
Given a Nash distribution $\vec{\pi}_L$ with \textbf{modification-L}, the following inequality holds for any policy $\pi^i$ and information set $x_t^i$ related to player $i$:
\begin{align*}
&\sum\limits_{a_{t}^{i}}{\pi _{L}^{i}(x_{t}^{i},a_{t}^{i})\sum\limits_{{{h}_{t}}\in x_{t}^{i}}{\Pr \left( {{h}_{t}}|{{{\vec{\pi }}}_{L}} \right)A_{{{{\vec{\pi }}}_{L}}}^{i}({{h}_{t}},a_{t}^{i})}}-\epsilon H(\pi _{L}^{i}(x_{t}^{i}))\sum\limits_{{{h}_{t}}\in x_{t}^{i}}{\Pr \left( {{h}_{t}}|{{{\vec{\pi }}}_{L}} \right)}\\
\geq\ &\sum\limits_{a_{t}^{i}}{{{\pi }^{i}}(x_{t}^{i},a_{t}^{i})\sum\limits_{{{h}_{t}}\in x_{t}^{i}}{\Pr \left( {{h}_{t}}|{{{\vec{\pi }}}_{L}} \right)A_{{{{\vec{\pi }}}_{L}}}^{i}({{h}_{t}},a_{t}^{i})}}-\epsilon H({{\pi }^{i}}(x_{t}^{i}))\sum\limits_{{{h}_{t}}\in x_{t}^{i}}{\Pr \left( {{h}_{t}}|{{{\vec{\pi }}}_{L}} \right)}
\end{align*}
\end{lemma}

\begin{lemma}
Given a Nash distribution $\vec{\pi}_R$ with \textbf{modification-R}, the following inequality holds for any policy $\pi^i$ and information set $x_t^i$ related to player $i$:
\begin{align*}
&\sum\limits_{a_{t}^{i}}{\pi _{R}^{i}(x_{t}^{i},a_{t}^{i})\sum\limits_{{{h}_{t}}\in x_{t}^{i}}{\Pr \left( {{h}_{t}}|{{{\vec{\pi }}}_{R}} \right)A_{{{{\vec{\pi }}}_{R}}}^{i}({{h}_{t}},a_{t}^{i})}}-\epsilon H(\pi _{R}^{i}(x_{t}^{i}))\sum\limits_{{{h}_{t}}\in x_{t}^{i}}{\Pr \left( {{h}_{t}}|{{{\vec{\pi }}}_{R}} \right)}\\
\geq\ &\sum\limits_{a_{t}^{i}}{{{\pi }^{i}}(x_{t}^{i},a_{t}^{i})\sum\limits_{{{h}_{t}}\in x_{t}^{i}}{\Pr \left( {{h}_{t}}|{{{\vec{\pi }}}_{R}} \right)A_{{{{\vec{\pi }}}_{R}}}^{i}({{h}_{t}},a_{t}^{i})}}-\epsilon H({{\pi }^{i}}(x_{t}^{i}))\sum\limits_{{{h}_{t}}\in x_{t}^{i}}{\Pr \left( {{h}_{t}}|{{{\vec{\pi }}}_{R}} \right)}
\end{align*}
\end{lemma}

Formal proofs for \textbf{Lemma 2} and \textbf{Lemma 3} are provided in Appendix B.3 \& B.4.

With this inequality, we can instantly observe a relationship between the two variants of the Nash distribution and the rest points (specifically, fixed points with regard to $\vec{w}(\vec{\sigma}(\cdot))$) in IESL.

\begin{theorem}
Every Nash distribution $\vec{\pi}_*$ with either \textbf{modification-L} or \textbf{modification-R} induces a rest point $\vec{y}_*=\vec{w}(\vec{\pi}_*)$ in IESL.
\end{theorem}

Although the proof for \textbf{Theorem 1} can be completed directly using the definition of IESL, we still leave it in Appendix B.5 to save space.

While the two variants of Nash distribution yield approximate Nash equilibria and IESL rest points by definition, we need a further assumption on the concavity of the game to bound the deviation of the policies of the IESL rest points. By introducing the concept of local hypomonotonicity, we will also show how IESL yields convergent policies with bounded deviation in both theory (see Section 4) and practice (see Section 5).

\section{Convergence Analysis}

\subsection{Deviation Bound for IESL Rest Point}

Concavity (or convexity) is a basic assumption commonly used in game theory (see \cite{even2009convergence}\cite{bravo2018bandit}) and general optimization (see \cite{boyd2004convex}). Now we first show that under the concavity of the individual utility functions, every rest point of IESL induces an approximate Nash equilibrium.

\begin{theorem}
Assume that for any player $i$ and any fixed $\vec{\pi}^{-i}$, the individual utility function $u^i(\pi^i,\vec{\pi}^{-i})$ is concave over $\Pi^i$. Then every rest point $\vec{y}_\epsilon$ in IESL induces an $\epsilon^\prime$-Nash equilibrium $\vec{\pi}_\epsilon=\vec{\sigma}(\vec{y}_\epsilon)$, where $\epsilon^\prime=\epsilon \log \left| A \right|\sum\limits_{k=0}^{T-1}{{{\gamma }^{k}}}$.
\end{theorem}

A formal proof for \textbf{Theorem 2} is provided in Appendix B.6.

Similar to the variants of the Nash distribution, the deviation of the IESL rest points is bounded by ${{\epsilon }^{\prime }}=\epsilon \log \left| A \right|\sum\nolimits_{k=0}^{T-1}{{{\gamma }^{k}}}$, which is at most $\epsilon T\log\left|A\right|$ since $\gamma\in(0,1]$. To further demonstrate that the IESL dynamic yields policies converging to approximate Nash equilibria, we only need to show that the policy in IESL converges to the corresponding policy at a rest point.

\subsection{Convergence of IESL Dynamic}

Attributed to the policy-entropy regularization term in the choice map $\vec{\sigma}(\cdot)$ (\ref{sigma-def}), the convergence of IESL only requires certain assumptions on the concept of local hypomonotonicity. Appendix A.4 compares the other definitions related to game hypomonotonicity and shows that the restriction below is the minimum among them.

\begin{definition}
We say the game is \textbf{locally} $\bm{\mu}$\textbf{-hypomonotone about }$\bm{\pi_\dagger}$ ($\mu\geq0$) on a policy set $\Pi_{local}$ containing $\pi_\dagger$, if the following inequality holds for any $\vec{\pi}\in\Pi_{local}$:
\begin{align*}
\sum\limits_{i\in N}{\left\langle {{\pi }^{i}}-\pi _{\dagger }^{i},{{w}^{i}}(\vec{\pi})-{{w}^{i}}(\vec{\pi}_{\dagger }) \right\rangle }\le \mu \sum\limits_{i\in N}{{{\left\| {{\pi }^{i}}-\pi _{\dagger }^{i} \right\|}^{2}}}
\end{align*}
where the inner product sums over all $(k,h_k,a_k^i)$ triples.
\end{definition}

In \cite{mertikopoulos2019learning}, Mertikopoulos et al. examined a concept called Fenchel coupling, whose Lyapunov properties are useful in convergence analysis of equilibrium-learning dynamics based on duality (regarding policy space as primal space and score space as dual space). Since Fenchel coupling can be lower-bounded with the policy distance when using the strongly convex penalty function $H(\cdot)$, we can use this concept to prove local convergence of (\ref{IESL}) in the context of imperfect information games. To be specific, we can construct an energy function and demonstrate its non-increasing property under sufficiently strong local hypomonotonicity.

\begin{theorem}
Assume that a rest point $\vec{y}_\epsilon$ of IESL is contained in a positively invariant compact set $\Omega$ that induces a local policy set ${{\Pi }_{local}}=\left\{ \vec{\sigma }(\vec{y})\left| \ \vec{y}\in \Omega  \right. \right\}$, and local $\mu$-hypomonotonicity about $\vec{\pi}_\epsilon=\vec{\sigma}(\vec{y}_\epsilon)$ is satisfied on $\Pi_{local}$. When $\mu<\epsilon$, policy yielded by IESL starting from an arbitrary $\vec{y}_0\in\Omega$ converges to $\vec{\pi}_\epsilon$.
\end{theorem}

A formal proof for \textbf{Theorem 3} is provided in Appendix B.7.

For ease of understanding, Figure \ref{Figure 1} reviews the relationships among the concepts involved in our theoretical analysis, showing the rationale behind the construction of the IESL dynamic.

\begin{figure}
\begin{center}
\includegraphics[width=1.0\linewidth]{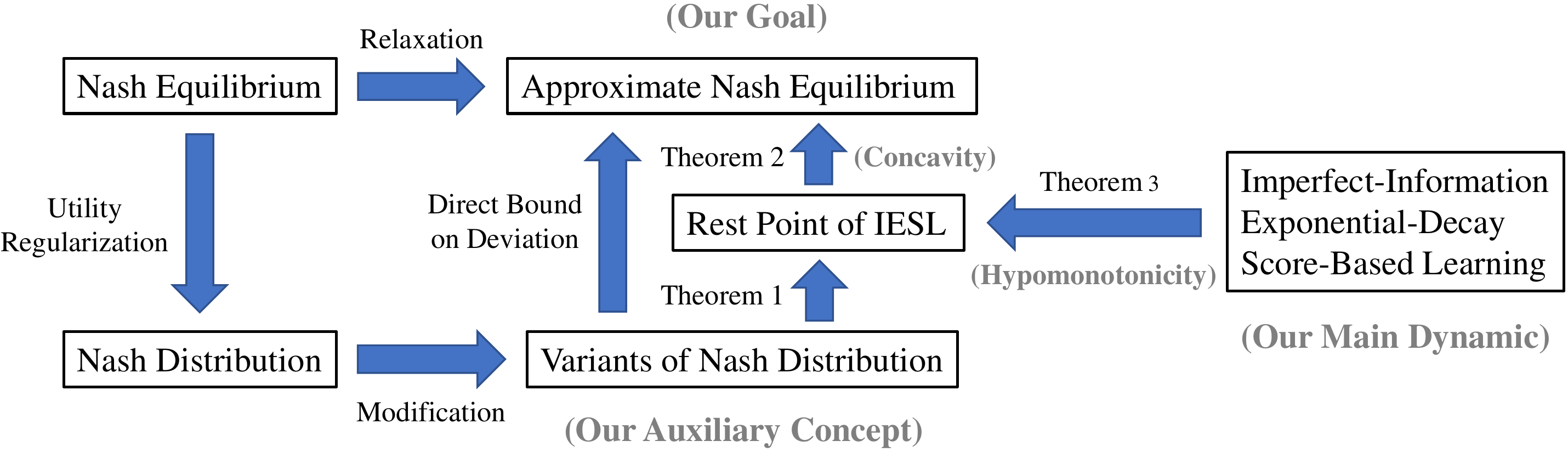}
\end{center}
\caption{Overview of our theoretical analysis}
\label{Figure 1}
\end{figure}

Note that \textbf{Theorem 3} suggests that whenever the score $\vec{y}$ in IESL falls into a region that is close enough to a rest point and satisfies certain local hypomonotonicity, the corresponding policy will inevitably converge to that of the rest point. Therefore, a trade-off exists with respect to the selection of $\epsilon$, since smaller $\epsilon$ guarantees lower NashConv for the policies that converge (under the concavity assumption in \textbf{Theorem 2}) but makes the dynamic convergence more difficult (as it requires $\mu<\epsilon$). An intuitive explanation for the influence of selecting different $\epsilon$ is provided in Appendix C.

In practice, using binary search as an efficient parameter selection method for $\epsilon$ in IESL, we are able to obtain policies approaching approximate Nash equilibria with competitively low NashConv. In Section 5, we will empirically show that our dynamic manages to find approximate Nash equilibria in multi-player poker scenarios and outperforms existing algorithms in moderate-scale Leduc poker games, demonstrating the equilibrium-finding capability of IESL even beyond theory (in practical sequential games).

\section{Empirical Results}

\subsection{Comparative Experiments}

\begin{wraptable}{r}{4.4cm}
\caption{Problem scale (\textbf{His.} means the number of histories and \textbf{Info.} means the number of information sets)}
\label{Table 1}
\begin{center}
\begin{tabular}{c|c|c}
& \textbf{His.} & \textbf{Info.}\\\hline
Kuhn-2 & 54 & 12\\
Kuhn-3 & 600 & 48\\
Leduc-2 & 9450 & 936\\
Leduc-3 & 396120 & 13878
\end{tabular}
\end{center}
\end{wraptable}

Since CFR \cite{zinkevich2007regret}, FP \cite{heinrich2015fictitious}, and RD \cite{hennes2020neural} are the major equilibrium-learning dynamics with theoretical guarantees in imperfect information games (though requiring average policy and the zero-sum condition), we seek to compare the empirical results of IESL with them. We chose four canonical poker game scenarios (2-player Kuhn, 3-player Kuhn, 2-player Leduc, and 3-player Leduc) in order to show the effectiveness of our dynamic in multi-player imperfect information games. To facilitate practical comparison, we first discretize the continuous time in IESL (\ref{IESL}) with iteration steps. While the theoretical analysis of IESL is based on POSG-like simultaneous games, we can simply adjust it into sequential games by computing the required advantage terms in the same way as the counterfactual values in CFR \cite{zinkevich2007regret}. As for the evaluation metric NashConv, it essentially requires best-response computation, which can be completed within linear time through tree-form dynamic programming in the same way as in extensive-form FP \cite{heinrich2015fictitious}.

\begin{wraptable}{r}{9.2cm}
\caption{NashConv of policies derived from the four dynamics}
\label{Table 2}
\begin{center}
\begin{tabular}{c|c|c|c|c}
& Kuhn-2 & Kuhn-3 & Leduc-2 & Leduc-3\\\hline
IESL (Ours) & 0.000245 & 0.000318 & \textbf{0.016365} & \textbf{0.052198}\\
CFR \cite{zinkevich2007regret} & 0.000240 & 0.000399 & 0.019648 & 0.466209\\
FP \cite{heinrich2015fictitious} & \textbf{0.000130} & \textbf{0.000067} & 0.113065 & 0.198394\\
RD \cite{hennes2020neural} & 0.000147 & 0.030282 & 0.047568 & 0.540433
\end{tabular}
\end{center}
\end{wraptable}

Table \ref{Table 1} shows the scale of the four poker game scenarios we selected as test environments. Table \ref{Table 2} shows the NashConv of policies derived from the four dynamics (instantaneous policy for IESL and average policy for CFR, FP, and RD). Adjustment to potential parameters is leveraged to reveal the near-optimal performance of each dynamic under the same number of iterations. FP performs excellently in simple Kuhn poker environments while struggling to close the NashConv gap as the problem scales. RD achieves a rather low NashConv in the two-player zero-sum scenarios, as is guaranteed in theory, but performs poorly when it comes to three players. Consistent with the theoretical results, IESL manages to find approximate Nash equilibria in all four poker game scenarios. As one of the most influential simple dynamics for solving imperfect information games, CFR performs closely to IESL in the three small-scale scenarios while failing to further reduce NashConv in the 3-player Leduc poker game (also observed in \cite{abou2010using}).

\subsection{Effect of \texorpdfstring{$\epsilon$}{ε} in IESL}

\begin{wrapfigure}{r}{9.5cm}
\begin{center}
\includegraphics[width=1.0\linewidth]{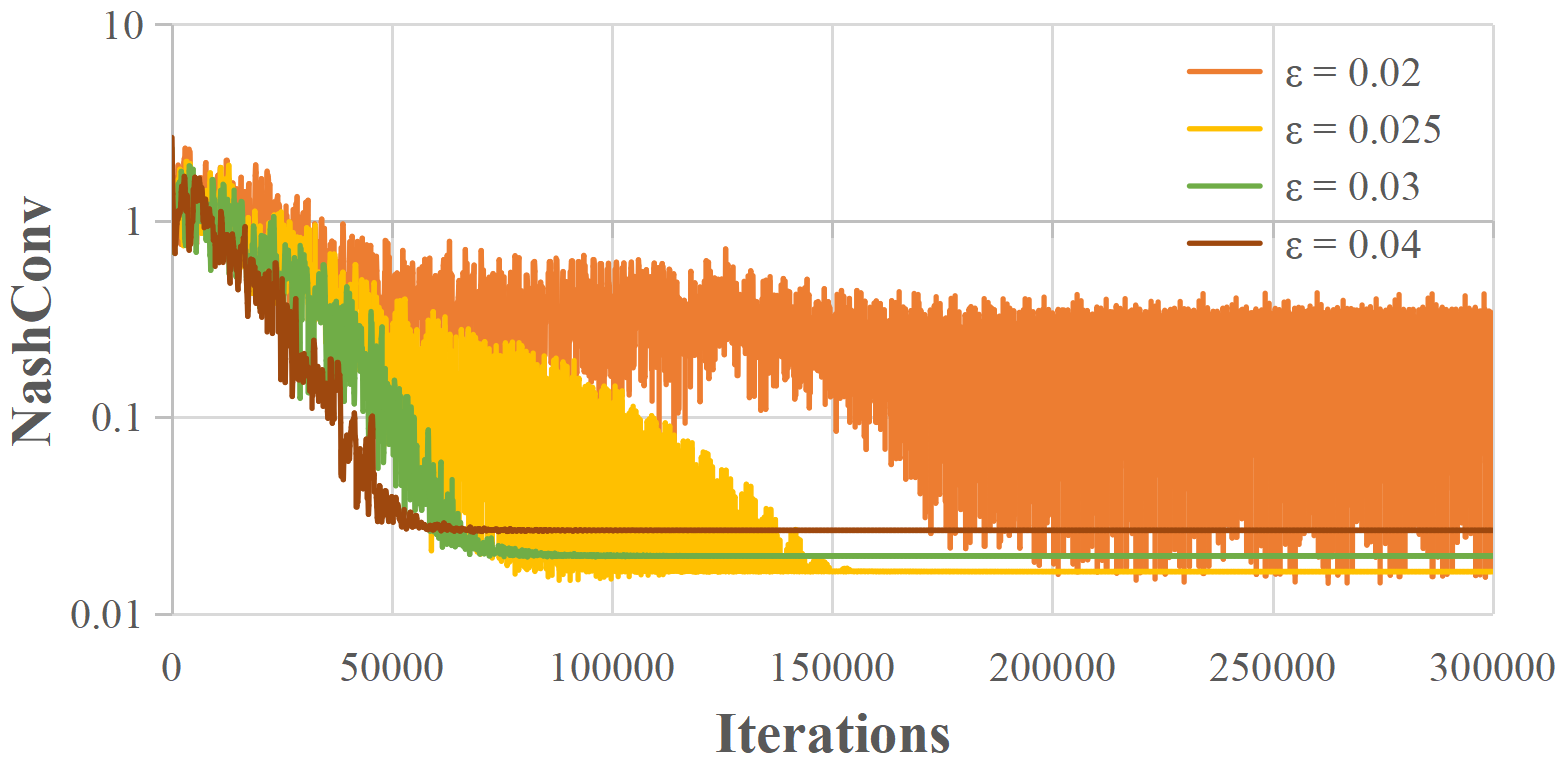}
\end{center}
\caption{Learning curves of IESL under different $\epsilon$}
\label{Figure 2}
\end{wrapfigure}

For IESL itself, it is worth noticing that a smaller $\epsilon$ leads to a less deviated converged policy but harder dynamic convergence, as is demonstrated in \textbf{Theorem 2} and \textbf{Theorem 3}. Figure \ref{Figure 2} compares the results of selecting different $\epsilon\in[0.02,0.04]$ in 2-player Leduc. While IESL finds rest points under $\epsilon \in \left\{ 0.04,0.03,0.025 \right\}$, it suffers from severe oscillation when we further reduce $\epsilon$ to $0.02$. For the three convergent curves, a small $\epsilon$ slows down the convergence speed but guarantees an ultimate policy with relatively low NashConv. This result is consistent with our theoretical analysis in Section 4, suggesting the generality of the theory even in games with real-world utility settings and non-simultaneous moves. Furthermore, when combined with theory, the result implies that the invariant areas restricted by local hypomonotonicity with $\mu<0.02$ around rest points (with regard to IESL under $\epsilon=0.02$) are significantly harder to find in the score space. This threshold phenomenon may reflect certain structures in the 2-player Leduc poker game.

In view of the results from both theory and experiment, an efficient method of optimal parameter selection is binary search. For example, given an initial interval of $(0.02,0.04]$ (with $\epsilon=0.02$ diverging and $\epsilon=0.04$ converging), we can first divide the interval into $(0.02,0.03]$ by checking the convergence under $\epsilon=0.03$, and further subdivide it into $(0.02,0.025]$ by checking the convergence under $\epsilon=0.025$. With a precision threshold of $0.005$, we derive the optimal parameter $\epsilon=0.025$ among $(0.02,0.04]$. For larger intervals like $(0,1]$, while the number of all possible $\epsilon$ is $\frac{1-0}{0.005}=200$, we can find the optimal $\epsilon$ within $\left\lceil{{\log }_{2}200}\right\rceil=8$ repetitions of IESL when using binary search to balance between policy convergence and ultimate NashConv.

More detailed information on experiments is provided in Appendix D, including the pseudocode of IESL implemented in sequential games (D.1), discussions on time complexity and learning curves (D.2), and a simple comparison with DRL algorithms (D.3).

\section{Conclusion}

\subsection{Overview}

In this paper, we covered three practical aspects not fully addressed in the current studies on imperfect information games: multiple players, simultaneous moves, and convergent instantaneous policies. By restating POSG as a general framework and proposing a corresponding learning dynamic, IESL, we examined the relationship among different classes of rest points in imperfect information games and demonstrated the equilibrium-finding capability of our simple dynamic, both theoretically and empirically. While computing approximate Nash equilibrium is generally hard (PPAD-complete), we showed that an efficient learning algorithm is possible when further restricting games with properties like local hypomonotonicity, whose existence seems not to contradict the experimental phenomena in practical poker games. The theoretical foundation established in this paper may ultimately facilitate real-life imperfect information game solving, which requires adaptability under a more general class of game rules. Besides, the manner in which our dynamic is constructed may also inspire researchers to propose new methods for dealing with multi-player games.

\subsection{Limitation \& Future Work}

A limitation of this work is that the IESL dynamic cannot be directly applied to solving large-scale games, as it requires traversing the entire game tree (expensive in time) and recording values and policies for all possible trajectories in tabular form (expensive in space). Therefore, our next research step is to construct a DRL extension for IESL in a similar way as Deep CFR \cite{brown2019deep} (from CFR), NFSP \cite{heinrich2016deep} (from FP), and NeuRD \cite{hennes2020neural} (from RD). Since the game description is compatible with the MDP nature, it is expected to be smooth to extend our theory to a DRL implementation with function approximation and randomization. Besides, since IESL shows a comparatively strong convergence result with regard to instantaneous policy in moderate-scale poker games, it is highly possible that an efficient implementation of equilibrium learning exists when it comes to large scale, with the help of neural networks and RL techniques like generalized advantage estimation (see \cite{schulman2015high}).

\bibliographystyle{plain}
\bibliography{references}

\end{document}